# Heterojunction Hybrid Devices from Vapor Phase Grown MoS$_2$

Chanyoung Yim[1,2], Maria O'Brien[1,2], Niall McEvoy[2], Sarah Riazimehr[3], Heiko Schäfer-Eberwein[3], Andreas Bablich[3], Ravinder Pawar[4], Giuseppe Iannaccone[4], Clive Downing[2], Gianluca Fiori[4], Max C. Lemme[3] and Georg S. Duesberg[1,2,*]

[1]School of Chemistry, Trinity College Dublin, Dublin 2, Ireland

[2]Centre for Research on Adaptive Nanostructures and Nanodevices (CRANN) and Advanced Materials and BioEngineering Research (AMBER) Centre, Trinity College Dublin, Dublin 2, Ireland

[3]University of Siegen, Hölderlinstrasse 3, 57076 Siegen, Germany

[4]Dipartimento di Ingegneria dell'Informazione, Università di Pisa, Via G. Caruso 16, 56122 Pisa, Italy

[*]Corresponding author: duesberg@tcd.ie

**Abstract**

We investigate a vertically-stacked hybrid photodiode consisting of a thin n-type molybdenum disulfide (MoS$_2$) layer transferred onto p-type silicon. The fabrication is scalable as the MoS$_2$ is grown by a controlled and tunable vapor phase sulfurization process. The obtained large-scale p-n heterojunction diodes exhibit notable photoconductivity which can be tuned by modifying the thickness of the MoS$_2$ layer. The diodes have a broad spectral response due to direct and indirect band transitions of the nanoscale MoS$_2$. Further, we observe a blue-shift of the spectral response into the visible range. The results are a significant step towards scalable fabrication of vertical devices from two-dimensional materials and constitute a new paradigm for materials engineering.

**Main**

Molybdenum disulfide ($MoS_2$), a semiconducting transition metal dichalcogenide (TMD), has drawn a lot of attention owing to its fascinating electronic and optical/optoelectronic properties.[1-4] Layered TMDs have a stacked two-dimensional (2D) lattice structure composed of an atomic plane of metal atoms sandwiched between two planes of chalcogen atoms. Recent studies on $MoS_2$ thin film field effect transistors have shown that the $MoS_2$ channel has very clear n-type characteristics with high mobility and good current on/off ratios,[5-7] which indicate that this layered material could be utilized in integrated circuits and logic circuit applications.[8,9] Further, it has been shown that monolayer $MoS_2$ has a direct band gap of ~1.8 eV, whereas bulk $MoS_2$ has an indirect band gap of ~1.3 eV.[2,10-12] The ability to tune the optical band gap of $MoS_2$ by thickness modulation suggests a wide range of applications in optoelectronic devices including phototransistors and photodetectors.[10,13-15].

Even though the optoelectronic properties of $MoS_2$ were the subject of intense research in the late 1960s,[16] scalable and reliable device production remains challenging, with most recent (opto-)electronic devices having been based on mechanically exfoliated $MoS_2$ with e-beam defined contacts in a 3-terminal field effect transistor (FET) configuration. These investigations focused on field-induced charge carrier movements, however, photoconductivity measurements at the junction between layered $MoS_2$ and conventional semiconducting substrates have rarely been conducted, which may be due to the lateral size limit of mechanically exfoliated flakes.[17] Additionally, exfoliation techniques have limited reproducibility and scalability.[1,2,18-20] Recently, large-area growth techniques based on vapor phase sulfurization of thin Mo films have been adopted for the synthesis of $MoS_2$ thin films.[21,22] We have previously shown that vapor phase grown $MoS_2$ thin films are electrically viable and can be employed to make sensors showing ultra-high sensitivity to ammonia.[23]

In this work, we introduce p-n heterojunction diodes fabricated by transferring vapor phase grown n-type $MoS_2$ thin films onto p-type silicon (p-Si) substrates. Previously, we created high quality diodes by transferring monolayer graphene onto pre-patterned silicon substrates.[24,25] In a similar fashion, $MoS_2$ layers of varying thicknesses are transferred forming a vertical hybrid device. This design allows the $MoS_2$ film to be directly exposed to light. The effect of varying the incident light intensity, wavelength and $MoS_2$ film thickness was investigated. The devices reported here display extraordinary sensitivity to changes in illumination. The spectral response showed a very broad spectrum with contributions from indirect and direct band gap transitions. Further, we report an extension of the spectrum into the visible range.

$MoS_2$ thin films were synthesized by vapor phase sulfurization of Mo films of varying thickness. The thicknesses of the $MoS_2$ films were found to be 4.17 ±0.18, 8.26 ±0.29, 12.52 ± 0.26 and 15.96 ± 0.16 nm, respectively, by spectroscopic ellipsometry (SE)[26] (See the Supplementary Information for more details on the SE measurements). Raman spectra were used to assess the quality of the $MoS_2$ films, as shown in Figure 1(a). The spectra all show the characteristic signal of 2H-$MoS_2$ with no obvious contributions from carbon contamination, polymer residue or oxides. The positions of the $E^1_{2g}$ (~383 cm$^{-1}$) and $A_{1g}$ (~ 408 cm$^{-1}$) peaks of the $MoS_2$ films, which are related to in plane and out of plane vibrational modes, respectively, are labelled for clarity. Previous reports on mechanically exfoliated flakes have noted a divergence in these peaks for films progressing from few-layer (< 5) to bulk MoS$_2$.[12] In our case, all of the films resemble bulk MoS$_2$ except for the 4.17 nm film, which displays a blue-shifted $E^1_{2g}$ peak. The average thickness of 4.17 nm, as measured by spectroscopic ellipsometry, suggests an average number of layers of ~6. However, given that the films are polycrystalline, it is probable that the observed peak shift can be explained by the presence of some few-layer (< 5) crystals. For thicker films, the spectra are consistent over the entire film, indicating the homogeneity of the film synthesis (see Figure S1(a) – (d) of the Supplementary Information). High-resolution transmission electron microscopy (HRTEM) analysis of the 8.26 nm film transferred to a TEM grid is depicted in Figure 1(b), which indicates a polycrystalline structure in plane. Analysis of the selected area electron diffraction (SAED) pattern in various regions gave a Mo-Mo lattice spacing of ~0.32 nm which is in agreement with the literature value.[27-30] However, ~20 % of SAED results showed lattice spacing values between 0.306 – 0.308 nm, which is 4% less than the reported literature value of 0.32 nm (See Figure S2(a) – (b) of the Supplementary Information for more details on the SAED pattern analysis). We note that such an assignment is preliminary, even though the greatest care was taken in ensuring the accuracy of these measurements.

Large-scale MoS$_2$ films were transferred using a polymer support technique leaving the films mechanically and electrically intact. In order to fabricate photodiodes, MoS$_2$ films of ~1 cm$^2$ area were transferred onto the pre-patterned p-Si substrates, as described in the methods section. The native oxide layer on the exposed silicon surface was removed with HF prior to deposition, ensuring good electrical contact between the MoS$_2$ and the Si. As shown in Figure 2(a), one end of the transferred MoS$_2$ film was placed on the p-Si surface without touching the metal electrode on the p-Si substrate, while the other end was connected to the Au pad on the SiO$_2$ layer, forming an ohmic contact between them.[5,10,31] More details on the transfer process are presented in Figure S3(b) of the Supplementary Information.

A plot of current-voltage (*J-V*) measurements of the diode device with 12.52 nm thick MoS$_2$ is depicted in Figure 2(b). The MoS$_2$ layers were fully electrically intact and well contacted by the gold pad as detailed above and therefore current transport was dominated by the MoS$_2$/Si interface region. Clear rectifying behavior was observed in the dark. The forward *J-V* characteristics of a diode in dark conditions can be expressed using the diode equation,[32] the ideality factor ($n \geq 1$) of a diode, which represents how closely the diode follows ideal diode behavior and has a value of unity in the ideal case, can be extracted from it. Considering the effect of the series resistance of the device, which is an additional secondary resistance component observed in the high forward bias region of practical diode devices, it gives an ideality factor value of 1.68 with a series resistance value of 7.3 k$\Omega$, indicating good rectifying performance. Details of the diode parameter extraction are presented in the Supplementary Information.

The diodes exhibit obvious photoconductivity under illumination with a white light source as presented in Figure 2(b). While little variation in the current density is seen between dark and illuminated conditions under forward bias, there is an obvious distinction in the reverse bias region. In the dark the device is in the off state under reverse bias and there is low reverse leakage current, but while illuminated, an evident current increase is observed in the reverse bias region. A cross sectional view of the n-type MoS$_2$/p-Si diode structure and its energy band diagram in reverse bias under illumination are shown in Figure 3(a). The n-type MoS$_2$ and the p-Si substrate form a p-n heterojunction and the topside of the MoS$_2$ film is exposed to the light source. Upon the incidence of photons, the valence band electrons are excited to the conduction band, generating electron-hole pairs in the n-type MoS$_2$, depletion layer and p-type Si. In the depletion layer, the excited electrons and holes are accelerated to the n-type MoS$_2$ and to the p-Si, respectively, and they are then collected in the n-type MoS$_2$ and p-Si region. When the electrodes of the p-Si substrate and n-type MoS$_2$ thin film are connected to an external circuit, electrons will flow away from the n-type MoS$_2$ film to the p-Si side and holes will flow away from the p-Si substrates to the n-type MoS$_2$ side, generating a current.

Figure 3(b) and 3(c) show a *J-V* plot of the diode with the 12.52 nm thick MoS$_2$ film under various light intensities and a photocurrent density ($J_{ph}$) plot extracted from the *J-V* measurements in the reverse bias region. The incident light intensity was controlled by a solid state dimmer and expressed using a relative value for the full intensity of the white light source. When the light intensity increased from 0 to 100 %, the measured current density also showed an increasing trend in the reverse bias region proportional to the incident light intensity. From the photo response of the diode, which measures the output photocurrent for varying input light intensity, the relative responsivity of the diodes for the light can be compared. At reverse dc biases of V = -1 V and -2 V, the generated $J_{ph}$ shows a nearly linear increment giving a photocurrent value of 0.49 mA and 0.83 mA at the full intensity of the light source, respectively (Figure 3(d)). The larger responsivity at higher reverse dc biases can be attributed to the fact that the energy barrier height at the junction between n-type MoS$_2$ and p-Si increases due to the increasing external electric field when a higher reverse bias is applied to the junction. This means the electrical potential difference across the depletion layer at the junction becomes much larger, which results in a stronger acceleration of electrons and holes in the depletion region and therefore a higher current.

In addition, the effect of modifying the MoS$_2$ thickness on the photocurrent was investigated. The thickness values of MoS$_2$ films, as measured by SE, were found to be 4.17, 8.26 and 15.96 nm. The *J-V* plots exhibit clear photoconductivity for all the devices when they are illuminated by the light source, as shown in Figure S5(a) – (c) of the Supplementary Information. The $J_{ph}$ values of the diodes with different MoS$_2$ thicknesses under reverse bias are compared in Figure 3(e). It is clear from these results that when the same incident light intensity (50 % of full intensity) is applied, the diode device with a thicker MoS$_2$ thin film shows higher photocurrent values. This is because the volume of photon absorption in the n-type MoS$_2$ thin film becomes larger with increased MoS$_2$ film thickness. As more photons are absorbed in the thicker MoS$_2$ layer more electron-hole pairs are produced, increasing the photocurrent of the device. This implies that it is possible to define the level of photoconductivity for MoS$_2$/Si photodiode devices through modulation of the MoS$_2$ film thickness.

The absolute spectral response was measured using a lock-in technique with a chopped photon flux over an investigated wavelength interval of 10 nm. Since the collimated incident light beam is bigger than the probe active area, a mask was used to define an illumination spot of 20 mm$^2$ on the device, either on the MoS$_2$ region or on the p-Si next to the MoS$_2$ (Figure 4 (a)). This allows contributions from the blank p-Si substrate to the generated photocurrent to be eliminated. The spectral response of the diode device with an 8.26 nm thick MoS$_2$ film for a reverse bias voltage between 0 – 2 V is plotted in Figure 4 (b). The multilayer MoS$_2$ photodiode exhibits a wide spectral response, which increases with higher reverse bias voltages due to the increase of the external electrical field. The spectral responsivity of 1.4 – 8.6 mA/W at reverse dc biases of V = -2 V is achieved in the broad spectral range from visible to near-infrared. While this performance is inferior to that of recently reported monolayer (800 A/W) and multilayer MoS$_2$ (thickness 30 – 60 nm, 120 – 210 mA/W) photodetectors[14,33,34], it is better than previously reported monolayer MoS$_2$ phototransistors (~7.5 mA/W)[13] and graphene photodetectors (~0.5 mA/W)[35]. The underlying p-type silicon absorption peak is observed at approximately 1.07 eV (1158 nm) confirmed in a reference measurement where the illumination spot was moved to the p-Si (See Figure S6 in the Supplementary Information). Three additional peaks were identified through fitting in the spectrum at 1.43 eV, 2.15 eV and 2.48 eV. The first peak at 1.43 eV (867 nm) can be explained by the indirect band transition ($\Sigma_m$-$\Gamma_v$) of multilayer MoS$_2$ as indicated by the blue arrow in Figure 4(c). Moreover, there is a strong contribution from the direct band gap transition observed in nanoscale MoS$_2$ films. This is an interesting observation considering the thickness of the film is 8.26 nm and therefore approximately 12 layers thick. The direct band gap contribution is split into light- ($K_m$-$K_{v1}$) and heavy holes ($K_m$-$K_{v2}$). Therefore, two distinct peaks in the spectral response are observed at energies of 2.15 eV (576 nm) and 2.48 eV (500 nm), illustrated by the red and green arrow in Figure 4(c).

Interestingly, we observe a blue-shift of 0.13 eV for the indirect transition and of approximately 0.4 eV for both direct transitions compared to the theoretical values for exfoliated $MoS_2$ (i.e. 1.3 eV for the indirect band transition and 1.8 eV and 2.0 eV for light- and heavy holes for the direct band transition)[2,33] and those observed experimentally.[20,26] This blue shift in the spectral response of bulk $MoS_2$ compared with the absorption spectrum was previously mentioned by Wilson and Yoffe.[16]

Density functional theory (DFT) calculations were employed to investigate potential causes of the observed blue shift. In particular, the influence of both inter-layer spacing and lattice spacing on the band structure of multilayer $MoS_2$ was studied. DFT calculations were performed to compute the electronic band structure of bulk $MoS_2$. Computed bands are shown in Figure 4(c), where the maxima ($K_m$, $\Sigma_m$) and the minima ($K_{v1}$, $K_{v2}$, $\Gamma_v$) of the conduction and valence bands, respectively, are highlighted. In Figure 4(d), we show the variations of the direct ($K_m$-$K_{v1}$/$K_{v2}$) and indirect band gap ($\Sigma_m$-$\Gamma_v$) with respect to the equilibrium case, as a function of the inter-layer distance (expressed in %). As soon as the interlayer distance is increased, the indirect band gap increases, while the direct band gap is negligibly affected. This is in agreement with recent simulations performed on bilayer $MoS_2$.[15] $\Gamma_v$ are characterized by *p*-orbital wave-functions centered in correspondence of S atoms, while K minima are related to *d*-orbitals localized around Mo atoms. As a consequence, K points are less affected by increases in the interlayer distance, while the opposite holds for $\Gamma_v$ minimum. In Figure 4(e), we show the same quantities as above, but as a function of lattice spacing variation. In this case, we qualitatively reproduce the larger shift of the direct band gaps with respect to the indirect band gap as soon as the lattice is compressed. From a quantitative point of view, a compression of 4% serves to explain the observed blue-shifted peak positions for both the indirect and direct transitions in the spectral response analysis of the device. Thus we tentatively assign the observed blue-shifted absorption peaks to the disordered lattice of our $MoS_2$ films. This assumption is supported by our HRTEM analysis which shows slight compression in regions of our polycrystalline films. This could stem from the high temperature growth mechanism or could possibly be caused by the film transfer process. It must be noted here that a difference in the strain transfer and relaxation mechanism may exist when the $MoS_2$ films are transferred onto different substrates, such as a TEM grid and a silicon substrate. Further, we cannot completely rule out the presence of dopants and contamination (e.g. oxygen) in the films, which could lead to lattice distortions. Nonetheless, the variation of the photo-response of 2D TMDs is an exciting and unexpected finding and will undoubtedly be the subject of further investigation, particularly as this variation in the spectral response of layered TMDs could present a wide range of opportunities in material design.

In summary, p-n heterojunction diodes were fabricated using n-type $MoS_2$ films with varying thicknesses and p-Si substrates. $MoS_2$ thin films are laterally in contact with p-Si substrates over the substrate area, allowing for direct exposure to incident light of varying intensity. Electrical measurements revealed that the n-type $MoS_2$/p-Si diodes have good rectifying behavior as well as clear photoconductive characteristics. The photocurrent of the device has a strong dependence on the $MoS_2$ film thickness whereby the thicker $MoS_2$ films produce more photocurrent due to their increased volume for photon absorption. This demonstrates the potential to control the photocurrent of $MoS_2$/Si diodes by modulating the thickness of the $MoS_2$ layer. The spectral response of the device showed that there are contributions from direct and indirect band transitions in the multilayer MoS$_2$ film. We further observed a substantially extended spectral range for our device into the visible range.

By employing a polymer support transfer process for the MoS$_2$ thin films the MoS$_2$/Si hybrid structure, which combines a semiconducting nanoscale TMD and a traditional semiconducting material, was realized. This approach could potentially be extended to various other semiconducting materials in such hybrid structures. Further comprehensive studies of such structures are required to improve device performance and engineer the properties of the interface. Nevertheless, this type of hybrid device demonstrates the benefits of using long developed semiconducting technology to take advantage of the novel properties of nanomaterials for future nano- and optoelectronic devices.

**Methods**

Commercially available lightly doped p-Si wafers with a thermally grown silicon dioxide (SiO$_2$) layer (292 nm) were used as substrates. The p-Si wafer had a dopant (boron) concentration of $2.5 \times 10^{15}$ cm$^{-3}$ and <100> orientation. A part of the SiO$_2$ layer was completely etched by immersing it in 3 % diluted hydrofluoric acid (HF) for 20 minutes, resulting in a sloped sidewall at the Si/SiO$_2$ boundary. Using a shadow mask, nickel (Ni) and gold (Au) metal electrodes (Ni/Au = 20/50 nm) were deposited on top of the remaining SiO$_2$ layer and the exposed p-Si area. In order to achieve ohmic contacts between p-Si and Ni, the substrate was annealed at 400 ℃ under N$_2$ flow for 5 minutes.

MoS$_2$ thin films were synthesized using a vapor phase sulfurization process similar to the method described previously[22,23]. Mo films of varying thickness were deposited on SiO$_2$/Si substrates using a Gatan Precision Etching and Coating System (PECS), where the Mo film deposition rate (< 0.1 nm/s) and thickness were monitored using a quartz crystal microbalance. Sulfurization of the Mo samples took place in a quartz tube furnace consisting of two different heating zones. The sputtered films were placed in a zone which was heated to 750 °C and annealed for 30 minutes at a pressure of ~1 Torr under an argon (Ar) flow of 150 sccm (standard cubic centimeters per minute). Sulfur powder was heated to its melting point (113 °C) in the inlet zone of the furnace, and the generated sulfur vapor was supplied to the Mo films, where it reacted to produce $MoS_2$. This technique yielded a continuous multilayer of $MoS_2$. A schematic diagram of the process is shown in Figure S3(a) of the Supplementary Information.

The as-grown $MoS_2$ thin films were transferred onto various substrates. A polymer support technique was employed for the transfer process, whereby polymethyl methacrylate (PMMA, MicroChem) was spin-coated onto the $MoS_2$. The films were then floated on 2 M NaOH at room temperature until the $SiO_2$ layer between the $MoS_2$ and the Si substrates was completely etched away, leaving $MoS_2$/PMMA films floating on the surface. After cleaning in deionized water the films were transferred onto arbitrary substrates. The PMMA support layer was then dissolved in acetone at room temperature for 20 minutes.

Thickness measurements of $MoS_2$ thin films were carried out using a spectroscopic ellipsometry tool (Alpha SE, J. A. Woollam Co., Inc.) operating in the wavelength range of 380 – 900 nm at an angle of incidence of 70 °. HRTEM analysis was performed in an FEI Titan transmission electron microscope at an acceleration voltage of 300 kV. Diffraction patterns were acquired at a camera length of 580 mm to expand the 100 diffraction ring; this increased the pixel count and improved the accuracy of the lattice measurements. $MoS_2$ films were prepared for TEM characterization by floating the layers as described previously, then, from water dredging them onto a 300 mesh lacey carbon copper TEM grid (Agar Scientific). Raman spectra were obtained with a Witec Alpha 300 R confocal Raman microscope, using an excitation wavelength of 532 nm with a power of < 1 mW and a spectral grating with 1800 lines/mm. Electrical measurements were conducted on a Suss probe station connected to a Keithley 2612A source meter unit under ambient conditions. The metal electrode on the p-Si substrate was positively biased and the electrode on the $SiO_2$ layer was negatively biased. A white light source with a solid state dimmer for variable light intensity (ACE Light Source, SCHOTT: A20500, 150 watt halogen lamp) was used for photoconductivity measurements.

The spectral response was measured by a comparative method to a known spectral response of a reference detector using a Labview controlled setup. The light was generated by a tungsten-halogen and a deuterium-arc lamp, and covered the wavelength range of 200-2000 nm. Specific wavelengths were selected by a monochromator (Acton Research Corporation, SP-555) using appropriate grids and filters. The light power density varied from 1 µW/cm$^2$ at a wavelength of 200 nm up to 55 µW/cm$^2$ at a wavelength of 1150 nm. A silicon photodiode was used to calibrate the setup. Due to the spectral limitation of the silicon photodiode, the spectral response measurement was limited to the range of 400 nm – 1250 nm. The detector currents were measured by pre-amplifiers (Femto, DLPCA-200) and lock-in amplifiers (Princeton Applied Research Corporation, Model 5210) with 300 ms integration time and 0.4 Hz band width at 17 Hz optical chopper frequency for detection of ultra-low currents down to 10 pA. The measurement principle allows a wavelength dependent correction factor to be established for the probe spectral response calculation which takes into account variations of the preamplifiers, the difference between the reference detector area and the probe area as well as varying photo flux densities caused by the monochromator grids and filters.

Density functional theory (DFT) calculations were carried out to compute the electronic band structure of bulk $MoS_2$. A Local density approximation (LDA) was assumed, adopting the exchange-correlation function by Perdew et al..[36,37] An ultra-soft pseudo-potential description of the electron-electron interaction was used with valence electrons $4d^5$, $5s^1$ and $3s^2$, $3p^4$ of Mo and S atoms, respectively. An 80 Ry wave function and a 500 Ry charge density cut-off were taken into account. A Brillouin zone sampling was considered on a Gamma centered 11 × 11 × 2 grid. Geometry optimization was performed, relaxing both ions and lattices until the total energy variation was less than $10^{-6}$ eV. Once the optimized geometry was obtained, the band structure calculations were performed in order to gain insight into the role of inter-layer distance. All calculations were conducted exploiting the Quantum-ESPRESSO package.[38]

**Figures**

(a)

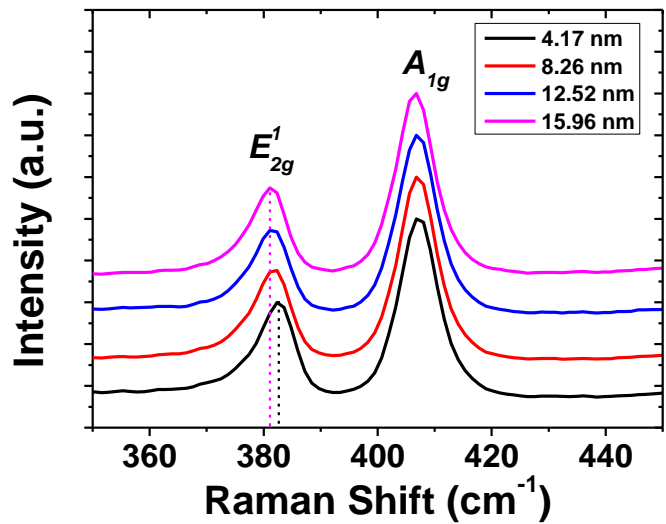

(b)

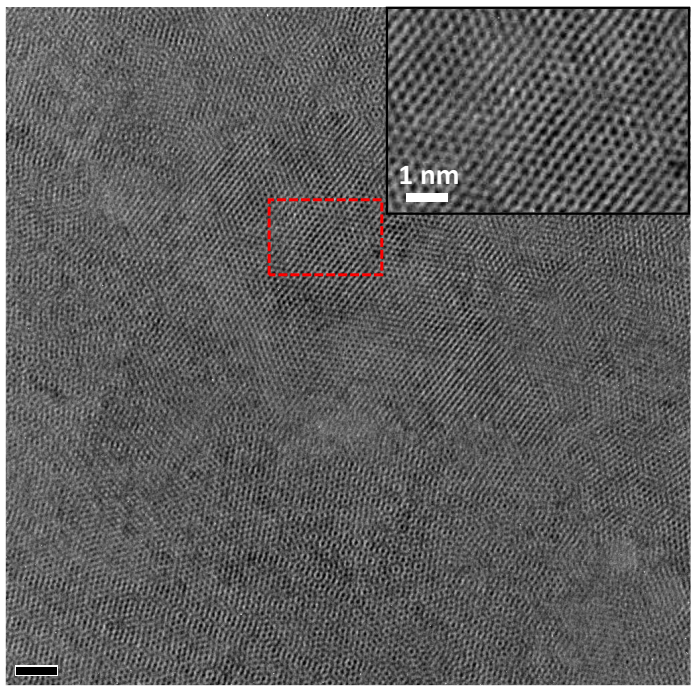

Figure 1. (a) Raman spectra of the MoS$_2$ thin films with various thicknesses grown by vapor phase sulfurization of Mo thin films. A slight shift of the $E^1_{2g}$ band is evident. (b) HRTEM image of a MoS$_2$ thin film transferred to a TEM grid (Inset: a corresponding image at high magnification).

(a)

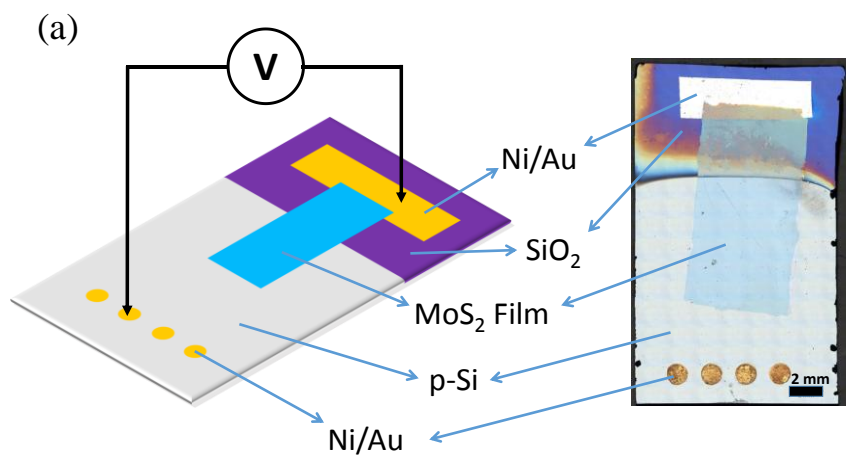

(b)

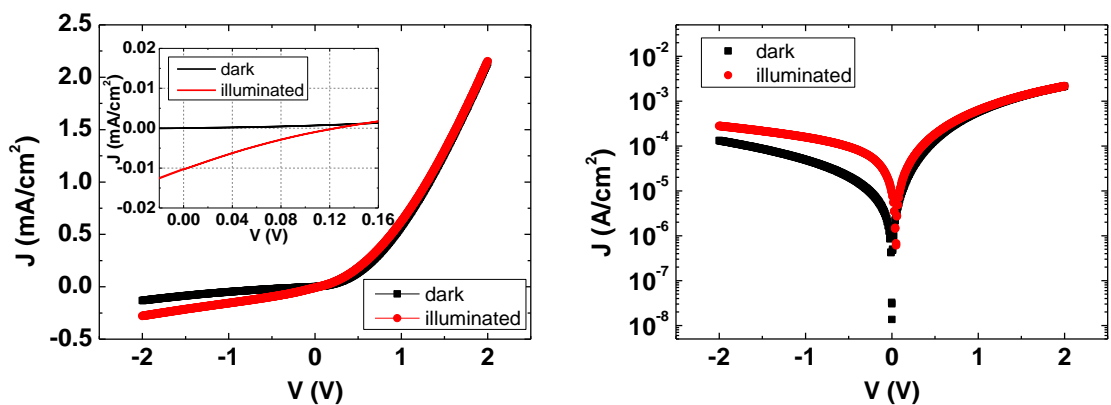

Figure 2. (a) Schematic (left) and photograph (right) of the n-type MoS$_2$/p-Si heterojunction diode. (b) A *J-V* plot of the diode with 12.52 nm thick MoS$_2$ on a linear scale (left) and semi-logarithmic scale (right) under dark (black) and illuminated (red) conditions. Inset of the left indicates open-circuit voltage (0.13 V) and short-circuit current (0.01 mA/cm$^2$).

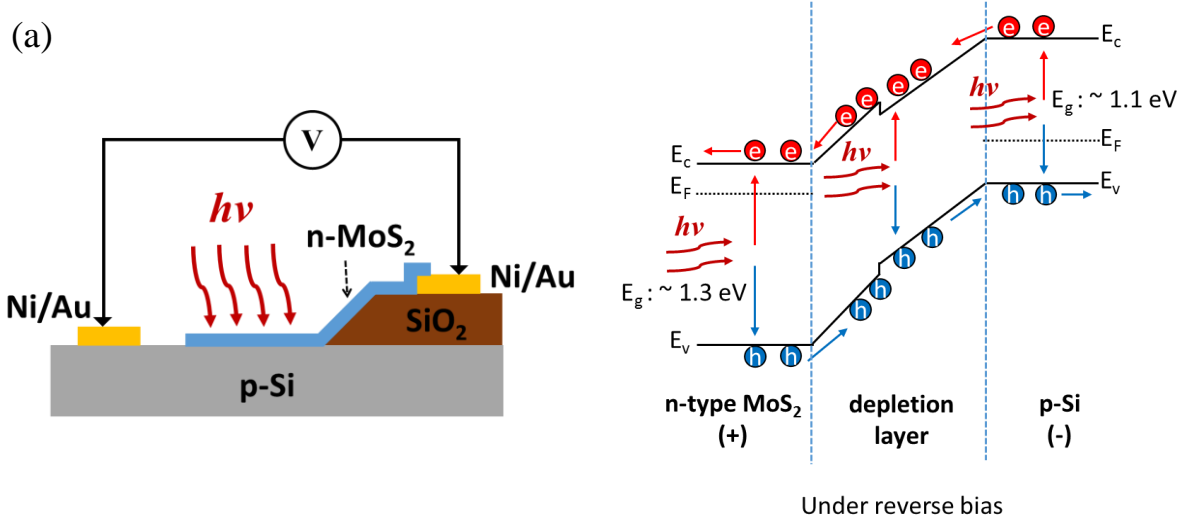

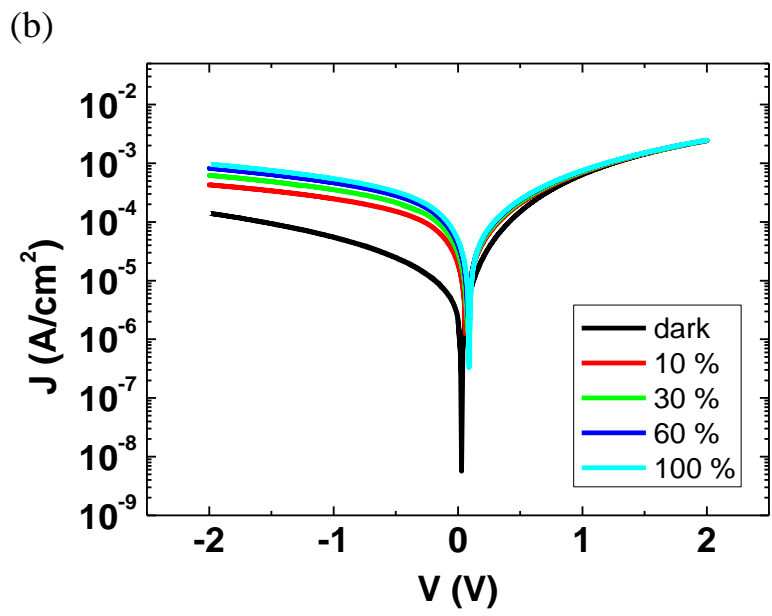

(c)

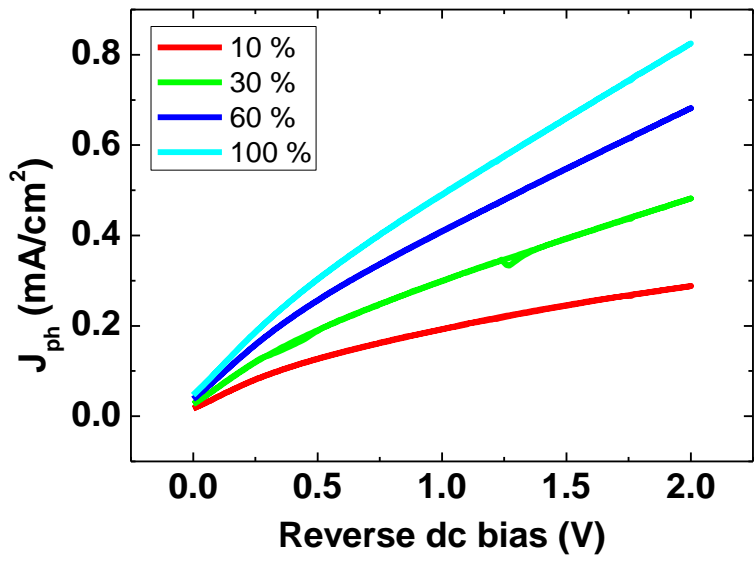

(d)

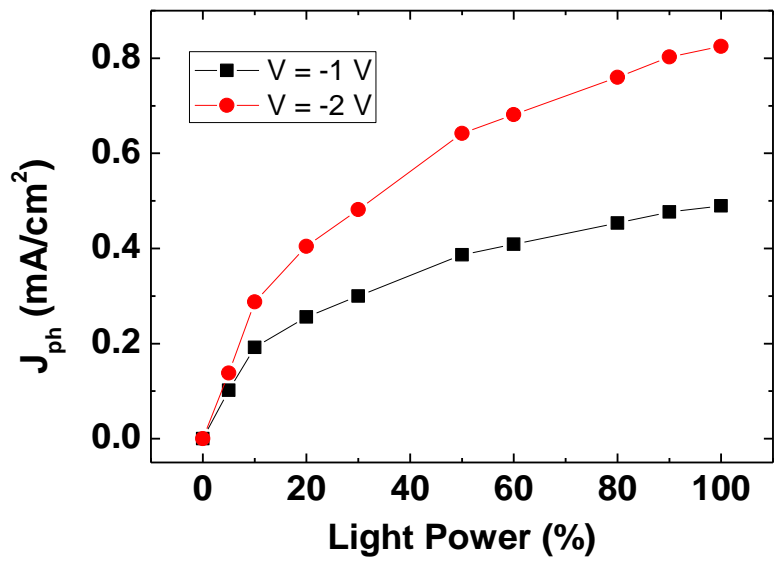

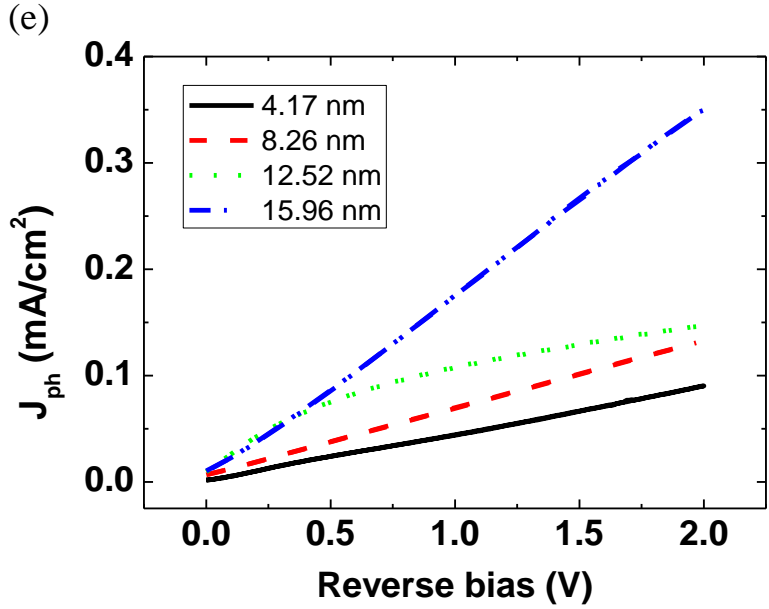

Figure 3. (a) Cross sectional view of the n-type $MoS_2$/p-Si diode structure (left) and its energy band diagram in reverse bias (right) under illumination, describing the movement of electrons (ⓔ) and holes (ⓗ). $E_c$, $E_F$, $E_v$, $E_g$ and $hv$ denote the conduction band, Fermi energy level, valence band, band gap and photon energy of the incident light, respectively. (b) *J-V* plot of the diode with the 12.52 nm thick $MoS_2$ film under various incident light intensities (dark, 10, 30, 60 and 100 % of full intensity) and (c) an associated photocurrent density ($J_{ph}$) plot extracted from the *J-V* measurements in the reverse bias region. (d) A $J_{ph}$ plot with varying incident light intensity at reverse biases of V = -1 and -2 V. (e) A plot of $J_{ph}$ of the diode devices with different $MoS_2$ thickness (4.17, 8.26, 12.52 and 15.96 nm) under reverse dc bias.

(a)

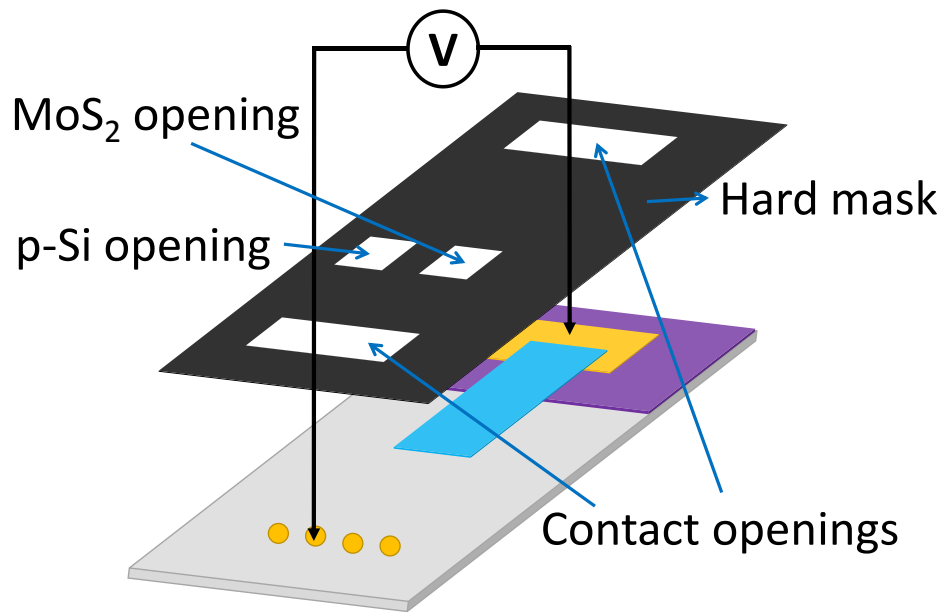

(b)

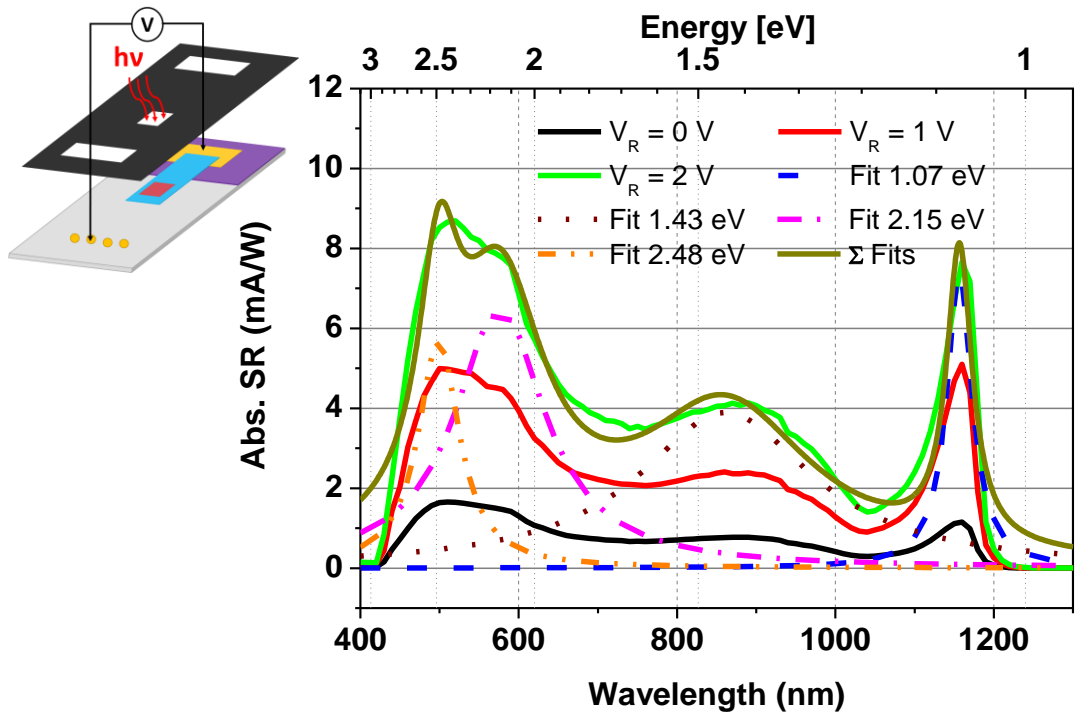

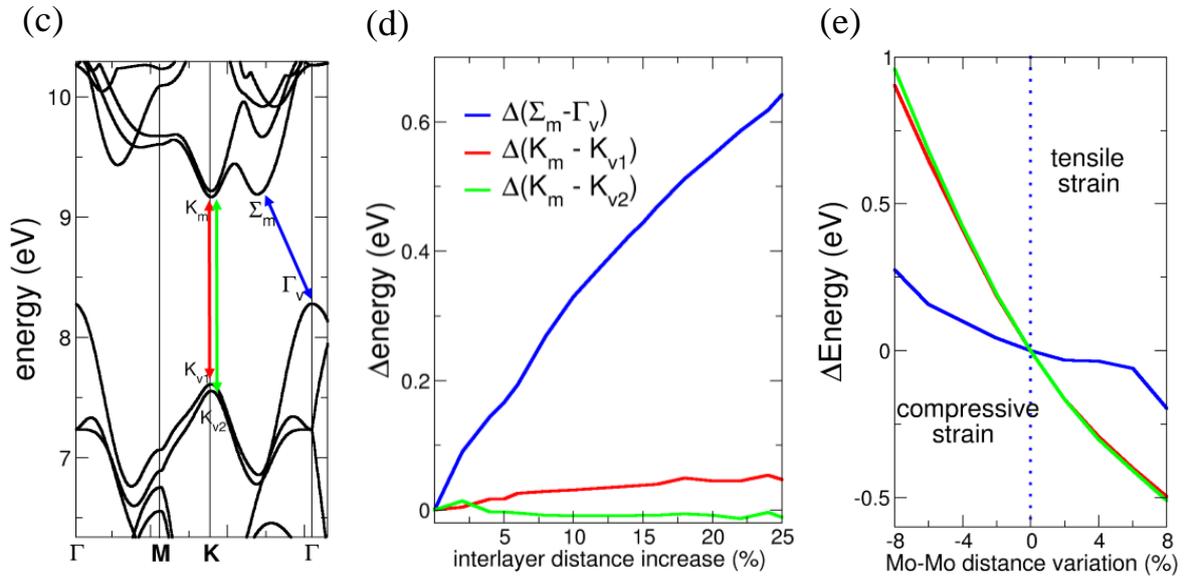

Figure 4. (a) Schematic of the heterojunction diode with mask openings for $MoS_2$ and p-Si indicated for spectral response measurements. (b) Absolute spectral response (Abs. SR) vs. wavelength (lower x-axis) and energy (upper x-axis) related to the diode device with an 8.26 nm thick $MoS_2$ film at zero bias and reverse bias ($V_R$) of 1 and 2 V with the mask opening on $MoS_2$. The inset indicates the illumination of the diode. (c) Calculated energy bands for bulk $MoS_2$. (d) Variation of the direct and indirect band gaps, with respect to the equilibrium case, as a function of the interlayer distance (expressed in %) and (e) variation of the direct and indirect band gaps, with respect to the equilibrium case, as a function of the lattice spacing (expressed in %).

Acknowledgements

This work is supported by the SFI under Contract No. 12/RC/2278 and PI_10/IN.1/I3030. C.Y. and M.O. acknowledge an Irish Research Council scholarship via the Enterprise Partnership Scheme. N.M. acknowledges the EU under FP7-2010-PPP Green Cars (Electrograph No. 266391). M.C.L. acknowledges financial support through an ERC grant (InteGraDe, No. 307311) as well as DFG (LE 2440/1-1 and GRK-1568). G.F. and G.I. gratefully acknowledge Quantavis s.r.l. through funds from FP7-GONEXTs project (Contract 309201). The authors thank the Advanced Microscopy Laboratory (AML) for their assistance in electron microscopy.

Author contributions

G.S.D. and C.Y. conceived and designed the experiments. M.O. and C.Y. fabricated the devices and performed the electrical measurements. C.Y. analyzed the data. N.M. carried out Raman spectroscopy measurements and analysis. M.O. and C.D. performed HRTEM analysis. S.R., H.S., A.B. and M.C.L. carried out the spectral response measurements and analysis. R.P., G.I. and G.F. performed the theoretical calculations. G.S.D. supervised the whole project. All authors contributed to the discussion of the results and improving the manuscript.

References

1   Novoselov, K. S. *et al.* Two-dimensional atomic crystals. *Proc. Natl. Acad. Sci. U. S. A.* **102**, 10451-10453 (2005).
2   Mak, K. F., Lee, C., Hone, J., Shan, J. & Heinz, T. F. Atomically Thin $MoS_2$: A New Direct-Gap Semiconductor. *Phys. Rev. Lett.* **105**, 136805 (2010).
3   Splendiani, A. *et al.* Emerging Photoluminescence in Monolayer $MoS_2$. *Nano Lett.* **10**, 1271-1275 (2010).
4   Korn, T., Heydrich, S., Hirmer, M., Schmutzler, J. & Schüller, C. Low-temperature photocarrier dynamics in monolayer $MoS_2$. *Appl. Phys. Lett.* **99**, 102109 (2011).
5   Radisavljevic, B., Radenovic, A., Brivio, J., Giacometti, V. & Kis, A. Single-layer $MoS_2$ transistors. *Nat. Nanotechnol.* **6**, 147-150 (2011).
6   Radisavljevic, B. & Kis, A. Mobility engineering and a metal–insulator transition in monolayer $MoS_2$. *Nat. Mater.* **12**, 815-820 (2013).
7   Wu, W. *et al.* High mobility and high on/off ratio field-effect transistors based on chemical vapor deposited single-crystal $MoS_2$ grains. *Appl. Phys. Lett.* **102**, 142106 (2013).
8   Radisavljevic, B., Whitwick, M. B. & Kis, A. Integrated Circuits and Logic Operations Based on Single-Layer $MoS_2$. *ACS Nano* **5**, 9934-9938 (2011).
9   Liu, L., Lu, Y. & Guo, J. On Monolayer $MoS_2$ Field-Effect Transistors at the Scaling Limit. *IEEE Trans. Electron Dev.* **60**, 4133-4139 (2013).
10  Lee, H. S. *et al.* $MoS_2$ Nanosheet Phototransistors with Thickness-Modulated Optical Energy Gap. *Nano Lett.* **12**, 3695-3700 (2012).
11  Schlaf, R., Lang, O., Pettenkofer, C. & Jaegermann, W. Band lineup of layered semiconductor heterointerfaces prepared by van der Waals epitaxy: Charge transfer correction term for the electron affinity rule. *J. Appl. Phys.* **85**, 2732-2753 (1999).
12  Lee, C. *et al.* Anomalous Lattice Vibrations of Single- and Few-Layer $MoS_2$. *ACS Nano* **4**, 2695-2700 (2010).
13  Yin, Z. *et al.* Single-Layer $MoS_2$ Phototransistors. *ACS Nano* **6**, 74-80 (2012).
14  Lopez-Sanchez, O., Lembke, D., Kayci, M., Radenovic, A. & Kis, A. Ultrasensitive photodetectors based on monolayer $MoS_2$. *Nat. Nanotechnol.* **8**, 497-501 (2013).
15  Zhao, W. *et al.* Origin of Indirect Optical Transitions in Few-Layer $MoS_2$, $WS_2$, and $WSe_2$. *Nano Lett.* **13**, 5627-5634 (2013).
16  Wilson, J. A. & Yoffe, A. D. The transition metal dichalcogenides discussion and interpretation of the observed optical, electrical and structural properties. *Adv. Phys.* **18**, 193-335 (1969).
17  Lopez-Sanchez, O. *et al.* Light Generation and Harvesting in a van der Waals Heterostructure. *ACS Nano*, **8**, 3042-3048 (2014).
18  Coleman, J. N. *et al.* Two-Dimensional Nanosheets Produced by Liquid Exfoliation of Layered Materials. *Science* **331**, 568-571 (2011).
19  Joensen, P., Frindt, R. F. & Morrison, S. R. Single-layer $MoS_2$. *Mater. Res. Bull.* **21**, 457-461 (1986).
20  Eda, G. *et al.* Photoluminescence from Chemically Exfoliated $MoS_2$. *Nano Lett.* **11**, 5111-5116 (2011).
21  Lee, Y.-H. *et al.* Synthesis of Large-Area $MoS_2$ Atomic Layers with Chemical Vapor Deposition. *Adv. Mater.* **24**, 2320-2325 (2012).
22  Zhan, Y., Liu, Z., Najmaei, S., Ajayan, P. M. & Lou, J. Large-Area Vapor-Phase Growth and Characterization of $MoS_2$ Atomic Layers on a $SiO_2$ Substrate. *Small* **8**, 966-971 (2012).
23  Lee, K., Gatensby, R., McEvoy, N., Hallam, T. & Duesberg, G. S. High Performance Sensors Based on Molybdenum Disulfide Thin Films. *Adv. Mater.* **25**, 6699-6702 (2013).
24  Kim, H.-Y., Lee, K., McEvoy, N., Yim, C. & Duesberg, G. S. Chemically Modulated Graphene Diodes. *Nano Lett.* **13**, 2182-2188 (2013).

25      Yim, C., McEvoy, N. & Duesberg, G. S. Characterization of graphene-silicon Schottky barrier diodes using impedance spectroscopy. *Appl. Phys. Lett.* **103**, 193106 (2013).
26      Yim, C. *et al.* Investigation of the Optical Properties of $MoS_2$ Thin Films Using Spectroscopic Ellipsometry. *Appl. Phys. Lett.* **104**, 103114 (2014).
27      Joensen, P., Crozier, E. D., Alberding, N. & Frindt, R. F. A study of single-layer and restacked $MoS_2$ by X-ray diffraction and X-ray absorption spectroscopy. *J. Phys. C: Solid State Phys.* **20**, 4043-4053 (1987).
28      Ataca, C. & Ciraci, S. Functionalization of Single-Layer $MoS_2$ Honeycomb Structures. *J. Phys. Chem. C* **115**, 13303-13311 (2011).
29      Ataca, C., Topsakal, M., Aktürk, E. & Ciraci, S. A Comparative Study of Lattice Dynamics of Three- and Two-Dimensional $MoS_2$. *J. Phys. Chem. C* **115**, 16354-16361 (2011).
30      Shi, Y. *et al.* van der Waals Epitaxy of $MoS_2$ Layers Using Graphene As Growth Templates. *Nano Lett.* **12**, 2784-2791 (2012).
31      Lembke, D. & Kis, A. Breakdown of High-Performance Monolayer $MoS_2$ Transistors. *ACS Nano* **6**, 10070-10075 (2012).
32      Neamen, D. A. *Semiconductor Physics and Devices : Basic Principles*. 3rd edn, 269-304 (McGraw-Hill, 2003).
33      Choi, W. *et al.* High-Detectivity Multilayer $MoS_2$ Phototransistors with Spectral Response from Ultraviolet to Infrared. *Adv. Mater.* **24**, 5832-5836 (2012).
34      Esmaeili-Rad, M. R. & Salahuddin, S. High Performance Molybdenum Disulfide Amorphous Silicon Heterojunction Photodetector. *Sci. Rep.* **3**, 2345 (2013).
35      Xia, F., Mueller, T., Lin, Y.-m., Valdes-Garcia, A. & Avouris, P. Ultrafast graphene photodetector. *Nat. Nanotechnol.* **4**, 839-843 (2009).
36      Perdew, J. P. *et al.* Atoms, molecules, solids, and surfaces: Applications of the generalized gradient approximation for exchange and correlation. *Phys. Rev. B* **46**, 6671-6687 (1992).
37      Perdew, J. P. & Wang, Y. Accurate and simple analytic representation of the electron-gas correlation energy. *Phys. Rev. B* **45**, 13244-13249 (1992).
38      Giannozzi, P. *et al.* QUANTUM ESPRESSO: a modular and open-source software project for quantum simulations of materials. *J. Phys.: Condens. Matter* **21**, 395502 (2009).

Supplementary Information

# Heterojunction Hybrid Devices from Vapor Phase Grown MoS$_2$

Chanyoung Yim[1,2], Maria O'Brien[1,2], Niall McEvoy[2], Sarah Riazimehr[3], Heiko Schäfer-Eberwein[3], Andreas Bablich[3], Ravinder Pawar[4], Giuseppe Iannaccone[4], Clive Downing[2], Gianluca Fiori[4], Max C. Lemme[3] and Georg S. Duesberg[1,2,*]

[1]School of Chemistry, Trinity College Dublin, Dublin 2, Ireland

[2]Centre for Research on Adaptive Nanostructures and Nanodevices (CRANN) and Advanced Materials and BioEngineering Research (AMBER) Centre, Trinity College Dublin, Dublin 2, Ireland

[3]University of Siegen, Hölderlinstrasse 3, 57076 Siegen, Germany

[4]Dipartimento di Ingegneria dell'Informazione, Università di Pisa, Via G. Caruso 16, 56122 Pisa, Italy

*Corresponding author: duesberg@tcd.ie

**MoS$_2$ thickness measurements using spectroscopic ellipsometry (SE)**

Thickness measurements were conducted using an Alpha SE tool (J. A. Woollam Co., Inc.) operating in the wavelength range of 380 - 900 nm at an angle of incidence of 70 ° with a beam spot size of ~40 mm$^2$. SE data were analyzed using CompleteEASE 4.72 (J. A. Woollam Co., Inc.). The SE system gathered values of psi ($\Psi$) and delta ($\Delta$) which represent the amplitude ratio ($\Psi$) and phase difference ($\Delta$) between p- and s-polarizations, respectively. The two parameters are related to the ratio $\rho$, defined by the equation of $\rho = r_p/r_s = \tan(\Psi)\exp(i\Delta)$, where $r_p$ and $r_s$ are the amplitude reflection coefficients for the p-polarized and s-polarized light, respectively.[1] A four-layer optical model which consists of a Si substrate, an interface layer between Si and SiO$_2$, a SiO$_2$ layer and a MoS$_2$ layer was built to analyze the SE spectra, and a Tauc-Lorentz (T-L) oscillation model[2] was used to determine the thicknesses of the MoS$_2$ thin films.

**Additional Raman analysis: Scanning Raman map**

Raman maps were acquired from the MoS$_2$ film (~12 nm thick) transferred onto SiO$_2$ at a power of 250 µW by taking 100 × 100 spectra over a 25 × 25 µm area (10, 000 discrete spectra). Each spectrum had an acquisition time of 0.2 s.

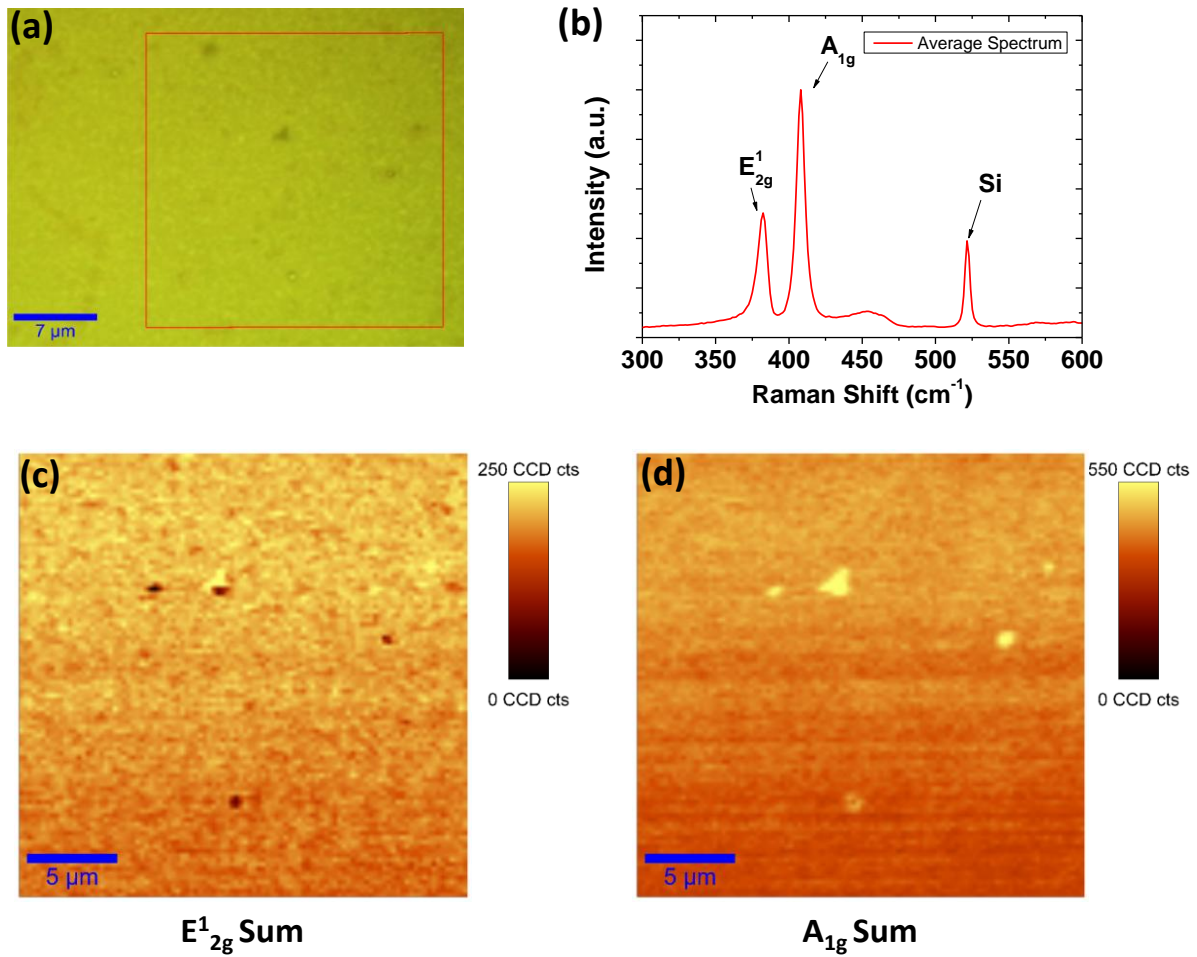

Figure S1. (a) Optical microscopy image of MoS$_2$ film transferred onto SiO$_2$. The red box indicates the 25 × 25 µm area over which Raman spectra were acquired. (b) Raman spectrum obtained by averaging over 10,000 discrete spectra which comprised the Raman map. Raman maps showing the intensity of the (c) $E^1_{2g}$ and (d) $A_{1g}$ peaks.

**Selected area electron diffraction (SAED) patterns of the MoS$_2$ thin film**

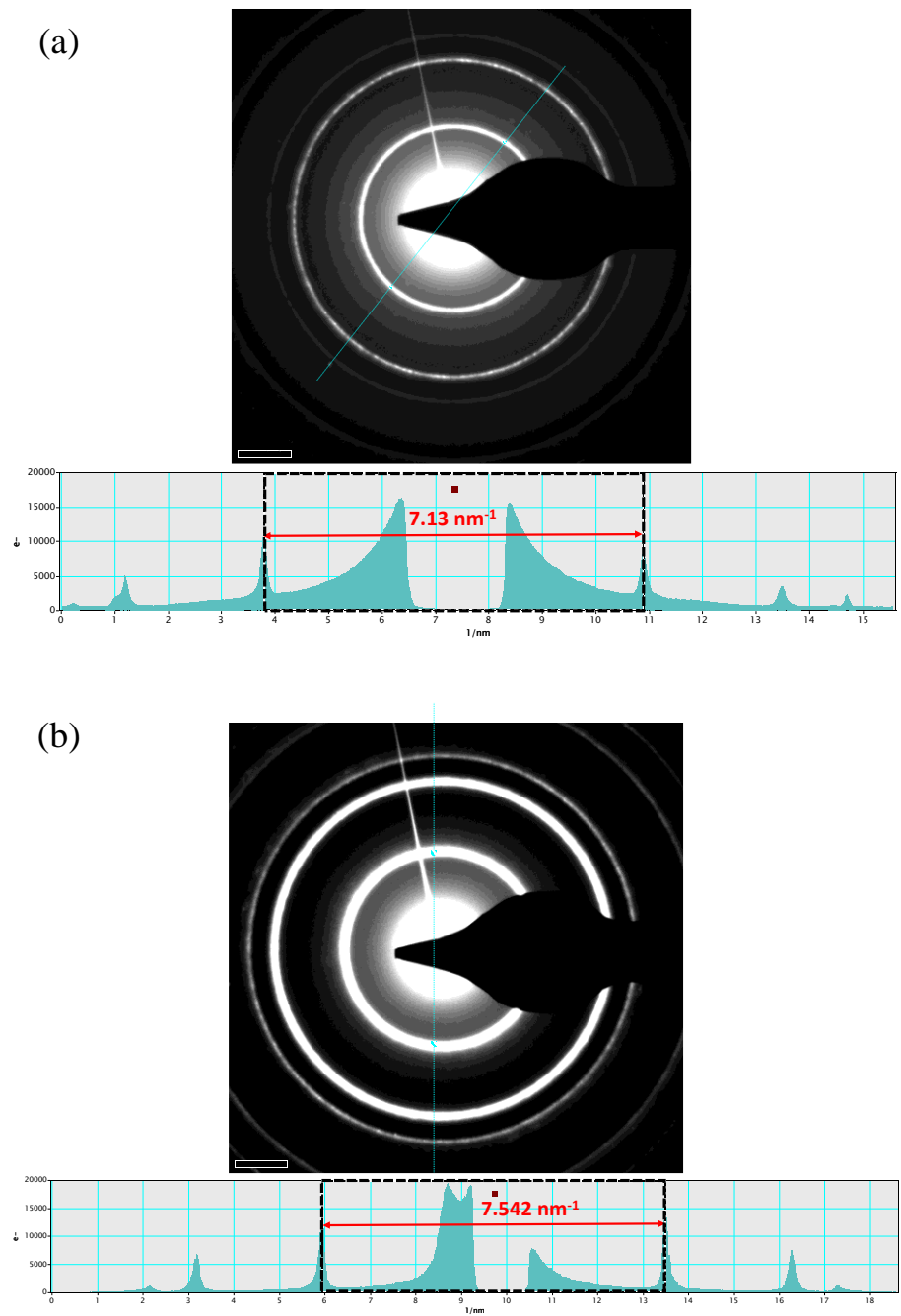

Figure S2. Representative electron diffraction patterns from different areas of the MoS$_2$ film revealing Mo-Mo lattice spacings of (a) 0.324 nm and (b) 0.306 nm, respectively.

**Device preparation**

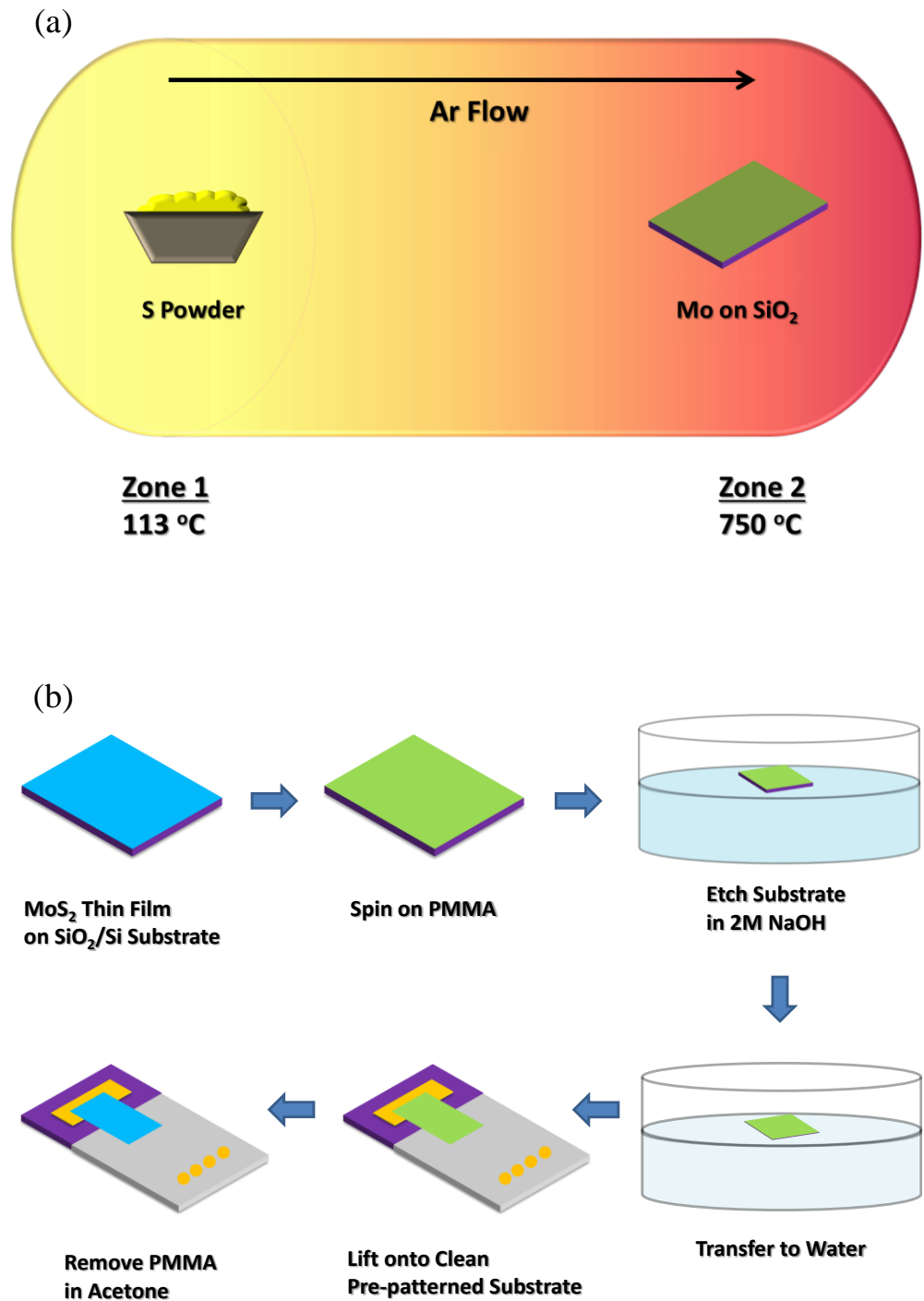

Figure S3. (a) Schematic diagram of the vapor phase sulfurization process. (b) Transfer process of as-grown MoS$_2$ thin films onto the pre-patterned substrate.

**Diode parameter extraction from the *J-V* measurements of the MoS$_2$/p-Si diode**

The forward *J-V* characteristics of a diode in dark conditions can be expressed using the flowing diode equation,

$$J = J_S[\exp\left(\frac{qV_D}{nk_BT}\right) - 1], \tag{1}$$

where *J* is the current density of the diode, $V_D$ is the voltage drop across the junction, $J_S$ is the reverse saturation current density, *n* is the ideality factor, $k_B$ is the Boltzmann constant, *q* is the elementary charge and *T* is the absolute temperature in Kelvin. When the effect of series resistance ($R_S$), an additional secondary resistance component observed in the high forward bias region of practical diode devices, is taken into account, for $V_D > 3k_BT/q$, by replacing $V_D$ with $V - JAR_S$, Eq (1) can be written as follows,

$$J = J_S \exp[\frac{q(V-JAR_S)}{nk_BT}], \tag{2}$$

where *V* is the total voltage drop across the series combination of the junction and $R_S$, and *A* is the effective contact area of the diode. This can be modified in terms of *V* and *J* and rewritten as the following equation.

$$V = JAR_S + \frac{nk_BT}{q}\ln\left(\frac{J}{J_S}\right), \tag{3}$$

Differentiating Eq. (3) in terms of *J*, we obtain

$$\frac{dV}{dJ} = AR_S + \frac{nk_BT}{qJ}. \tag{4}$$

According to Eq (4), *dV/dJ* is linearly proportional to *1/J*. Thus, the ideality factor ($n \geq 1$) of a diode, which represents how closely the diode follows ideal diode behavior and has a value of unity in the ideal case, and the series resistance, can be extracted by extrapolating the linear region of the plot of *dV/dJ* vs. *1/J*. The *dV/dJ* vs. *1/J* plot of the 12.52 nm device is shown in Figure S4, giving a series resistance value of 7.3 kΩ from the y-axis intercept and an ideality factor value of 1.68 from the slope.

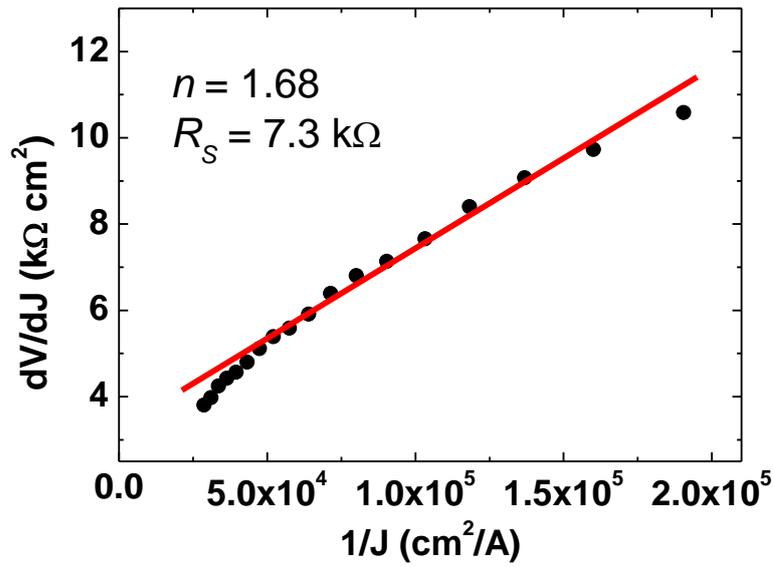

Figure S4. Plot of *dV/dJ* vs. *1/J* extracted from the *J-V* data of the diode with 12.52 nm thick MoS$_2$, giving a series resistance value of 7.3 kΩ and an ideality factor value of 1.68.

**Additional *J-V* plots of the MoS$_2$/p-Si diode devices**

(a)

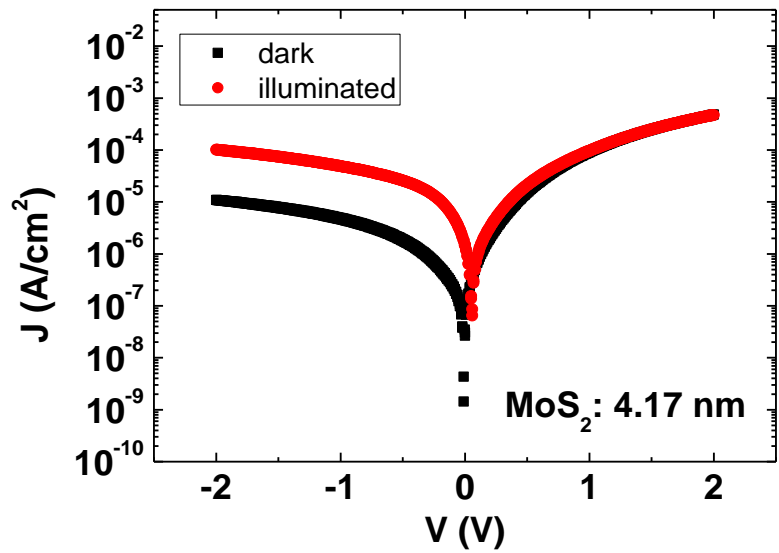

(b)

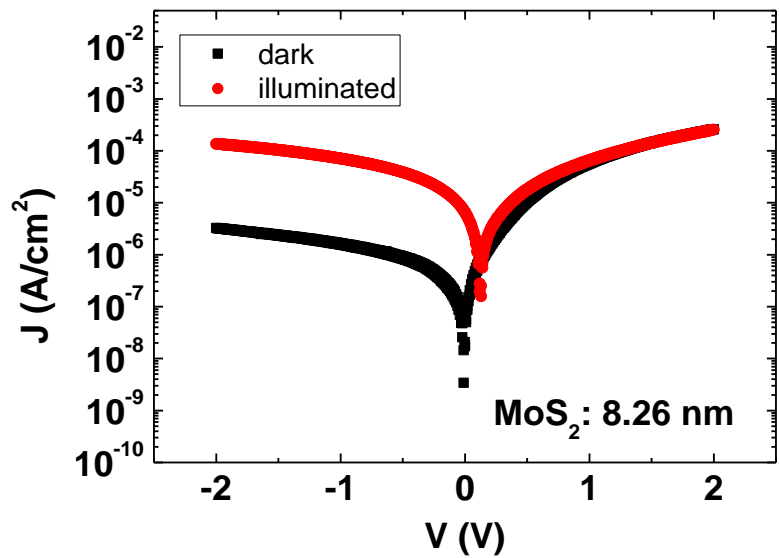

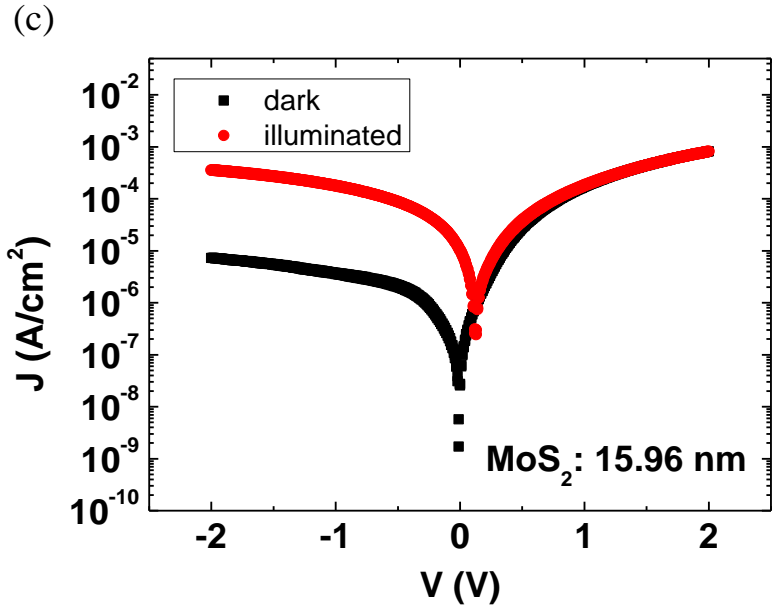

Figure S5. Semi-logarithmic *J-V* plots of the MoS$_2$/p-Si diode devices with MoS$_2$ film thickness of (a) 4.17 nm, (b) 8.26 nm, and (c) 15.96 nm in the dark and illumination (50 % of full intensity).

**Additional data for the spectral response measurements**

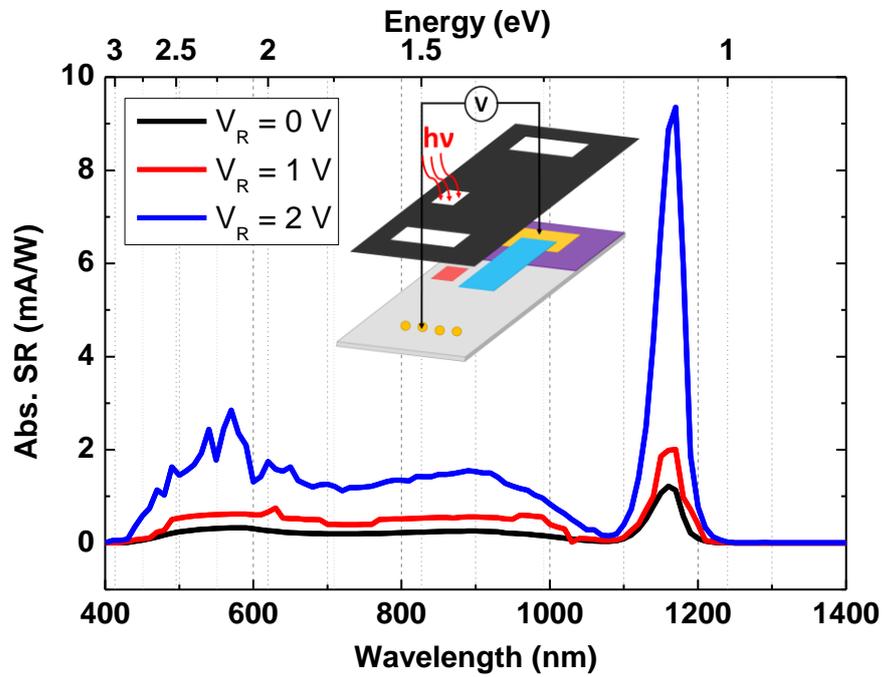

Figure S6. Absolute spectral response vs. wavelength (lower x-axis) and energy (upper x-axis) related to the diode device with an 8.26 nm thick $MoS_2$ film at zero bias and reverse bias ($V_R$) of 1 and 2 V with the mask opening on p-Si. The inner picture is a schematic diagram of the n-type $MoS_2$/p-Si heterojunction diode device with mask opening on p-Si.

References

1   Fujiwara, H. *Spectroscopic ellipsometry : principles and applications*.  81-87 (John Wiley & Sons Ltd, 2007).
2   Jellison, G. E. & Modine, F. A. Parameterization of the optical functions of amorphous materials in the interband region. *Appl. Phys. Lett.* **69**, 371 (1996).